\documentclass{aa}
\usepackage{amsmath}
\usepackage{natbib}
\bibpunct{(}{)}{;}{a}{}{,}
\usepackage{txfonts}
\usepackage{graphicx}
\usepackage{color}
\title{Intermittent heating in the solar corona employing a 3D MHD model}

\author{S.~Bingert\inst{1,2}\and H.~Peter\inst{1,2}}
\institute{Max Planck Institute for Solar System Research (MPS),
           37191 Katlenburg-Lindau, Germany, [bingert,peter]@mps.mpg.de
           \and
           Kiepenheuer-Institut f\"ur Sonennphysik (KIS),
           79104 Freiburg, Germany}
\date{Accepted 30 March 2011}
  \abstract
  {}
  {We investigate the spatial and temporal evolution of the heating of the
    corona of a cool star such as our Sun in a three-dimensional
    magneto-hydrodynamic (3D MHD) model.}
  {We solve the 3D MHD problem numerically in a box representing part of the
    (solar) corona. The energy balance
    includes Spitzer heat conduction along the
    magnetic field and optically thin radiative losses.  The self-consistent
    heating mechanism is based on the braiding of magnetic field lines
    rooted in the convective photosphere. Magnetic stress induced by
    photospheric motions leads to currents in the atmosphere that heat the
    corona through Ohmic dissipation.}
  {While the horizontally averaged quantities, such as heating rate,
    temperature, or density, are relatively constant in time, the simulated
    corona is highly variable and dynamic, on average reaching the temperatures
    and densities found in observations. The strongest heating per
    particle is found in the transition region from the chromosphere to the
    corona. The heating is concentrated in current sheets roughly aligned
    with the magnetic field and is transient in time and space. This
    supports the idea that numerous small heating events heat the corona,
    often referred to as nanoflares.}
  {}
\begin{document}
\maketitle
\section{Introduction}
The nature of the heating mechanism leading to a several million Kelvin hot
outer atmosphere of cool stars still remains elusive. It is generally agreed
that the mechanism heating the corona is related to the conversion from
magnetic to thermal energy. One fundamental problem is that the actual
dissipation of the (magnetic) energy will appear on microscopic scales,
while the observable structures in the corona are macroscopic.  A
comprehensive overview of coronal heating mechanisms can be found in the
classical conference proceedings of \cite{Ulmschneider+al:1991}.  For
example, if we assume that a magnetic resistivity follows from classical
transport theory, the dissipation should occur well below length scales of
1\,m. Likewise, the ion (electron) gyro-radii are close to m (cm) for
typical coronal magnetic fields, so that the ultimate dissipation process
has to operate on these small scales.

In contrast to this, the actually observed coronal structures on the real
Sun (and presumably on other stars) are on much larger scales. They range
from several Mm above magnetic patches of the chromospheric network to
hundreds of Mm for large active region loops, i.e. are almost close to
 the solar radius. It is clear that the gap in length scale to the
microscopic processes of a factor of some 10$^7$ to 10$^9$ cannot be bridged
by a model that aims to describe the whole range of coronal structures.

As a result, a large-scale model for actually observable solar structures has to
employ a (more or less sophisticated) parametrization of the heating
rate. \cite{Rosner+al:1978} solved static one-dimensional (1D) models for
coronal loops using different parametrization, including the assumption of a
(volumetric) heating rate that is constant in space and time, and derived their now
classical scaling laws relating coronal temperature and pressure to heating
rate and loop length. In a dynamic 1D loop model, \cite{Hansteen:1993}
assumed a heating rate that is transient in time and space by increasing the
internal energy from one time step to the next at one grid cell in the
numerical model, in order to study the response of the transition region to
nanoflare-like heating as originally proposed by
\cite{Parker:1988}. \cite{Mueller+al:2003,Mueller+al:2004} show that a
(volumetric) constant heating rate can lead to dynamic evolution and
catastrophic cooling within a coronal loop if the heating rate decreases
exponentially with height. \cite{Aiouaz+al:2005:model} investigated the
consequences of different parametrization of the heating rate on the outflow
profile from a coronal funnel.
\cite{2006ApJ...647.1452P} modeled a loop composed of many individual
(1D modeled) strands, each heated impulsively but constant in space. They
find significant deviations from a Gaussian line profile of the emergent
emission line, which still needs to be confirmed by
observations. \cite{Warren+al:2010} employ multi-stranded loop models, too,
by varying the energy input and the timing of the individual heating events in
order to match observed spatial and temporal variations of the emission from
coronal loops.
In a more global approach, \cite{Schrijver+al:2004} assume a parametrization
as a function of the magnetic field and loop length for (approximate) static
loop models. By comparing the appearance of the model coronae to the
observed structure, they infer the details of this parametrization.

This list of studies has to be incomplete, but all these (mostly 1D) models have in common that they assume the parametrization \emph{ad-hoc}.
By their very natur these models always suffer from ignoring the
spatial complexity and the changing 3D structure of the
magnetic field in which the loop under investigation is embedded. Of course
the major advantage of a 1D model is the high spatial resolution that can be
achieved and the possibility of including non-equilibrium ionization
\citep[e.g.,][]{Bradshaw+Mason:2003:basics,Bradshaw+Mason:2003:loop}.

A 3D magneto hydrodynamic (MHD) model can account for the
spatial complexity on the real Sun, and more important is that it allows a more
self-consistent treatment of how the heating rate is changing in space and
time. 
\cite{2010AAS...21630003M} use a parametrization depending on magnetic field
strength.
Through this they could compare the appearance of
the loops in the 3D model to 1D models.

The 3D MHD coronal models by
\cite{Gudiksen+Nordlund:2002,Gudiksen+Nordlund:2005a,Gudiksen+Nordlund:2005b}
include a heating based on field line braiding: foot point motions in
the photosphere deform the magnetic field, which induces currents that are
subsequently dissipated. This Ohmic heating is intermittent in time and
space as it depends on the evolution of the magnetic field which is
self-consistently modeled.
While the evolution of the magnetic field is treated self-consistently in
the 3D MHD model, the Ohmic heating rate, ${\eta}j^2$ should still be
considered as a parametrization. While the currents, $j$, are derived from
the magnetic field in the numerical models, the magnetic resistivity,
$\eta$, used in the model is much greater than the value derived from
transport theory \citep[e.g.][]{Boyd+Sanderson:2003}. This is due to
limitations in the magnetic Reynolds number in numerical
simulations. Therefore it is not clear whether ${\eta}j^2$ represents the true
heating rate.
Considering results of the models of \citeauthor{Gudiksen+Nordlund:2002} and
the good match of these models to solar spectroscopic observations
\citep{Peter+al:2004,Peter+al:2006}, it seems safe to at least consider
${\eta}j^2$ as a (good) parametrization of the actual heating rate.

This paper follows the model philosophy of
\cite{Gudiksen+Nordlund:2002,Gudiksen+Nordlund:2005a,Gudiksen+Nordlund:2005b}
and investigates the spatial and temporal variation of the heating rate in
an active region. Special attention is paid to the energy distribution of
individual energy releases, which can be considered as nanoflares.
The outline of the paper is as follows. In Sect.~\ref{sec:model-corona} we
describe our model and the set of MHD equations used. In
Sect.~\ref{sec:results} we analyze the data before we discuss the results
in Sect.~\ref{sec:discussion} and conclude in Sect.~\ref{sec:conclusion}.
\section{The model corona}
\label{sec:model-corona}
The numerical model includes the solar atmosphere above a small active
region in a 3D box of the size 50x50x30\,Mm, which corresponds
to roughly $0.05$ solar radii. The dynamics and heating of the corona 
stem from photospheric motions that braid the magnetic fields lines,
often called flux braiding. \footnote{Strictly speaking, \emph{flux}
  cannot be braided, thus this process should be called \emph{field line
    braiding.}} This field line braiding essentially depends on
  the field geometry and the driving boundary.
Successively currents are induced that heat the atmosphere by their
dissipation. More precisely the photospheric motions, together with the
magnetic field at high plasma beta, produce a Poynting flux into the upper
atmosphere.  The energy difference between the non-force-free field and a
potential field configuration, the free energy, is partly transferred to
heat via dissipation of currents. The time scale of the conversion is given
by the resistivity of the plasma.

The idea of heating by Ohmic dissipation has already been proposed by
\cite{1983ApJ...264..642P}. He estimated the energy flux above an active
region to be on the order of 100\,W\,m$^{-2}$ by computing the magnetic
stress introduced by the photospheric motions. The stress could be explained
by a strain of magnetic lines of forces, which are then rapidly dissipated by
reconnection. This paper follows that idea and investigates the rate and
the spatial distribution of the dissipation.

The temperature structure in the solar atmosphere is highly sensitive to the
radiative loss and the Spitzer heat conduction \citep{Spitzer:1962}. The
latter is proportional to $T^{5/2}$ and acts as a thermostat for the
corona. Energy input into the corona is conducted downwards along magnetic
lines of force into regions of higher density. With increasing density at
lower heights, the radiative loss becomes more efficient. If the heat input
into the corona is increased, more energy is transferred into the
chromosphere, and more energy has to be radiated away. Therefore the height
of the transition region where the Spitzer heat conduction brakes away moves
down to heights of higher densities to compensate for the increased energy
flux. As a result, the coronal density is higher when the atmosphere is in a
pressure equilibrium. But the average coronal temperature change is
small. Doubling the heating rate would only increase the temperature by a
factor of $2^{2/7}$ due to the efficient heat conduction.  In our model the
energy input by the Poynting flux at photospheric level is dissipated into
heat. Heat conduction transfers energy down to regions of high densities in
the lower atmosphere where it is radiated. We use the optically thin
radiative losses given by \cite{Cook+al:1989}.

The correct treatment of the energy equation is thus important for
estimating the position of the transition region, as well as temperature and
density of the corona. Direct comparisons with observations with synthesized
emission lines \citep{Peter+al:2004} are then possible. The intensity of the
optically thin coronal emission lines is proportional to the density
squared. Small changes in coronal densities directly influence the coronal
emissivities.  In the model of \cite{Gudiksen+Nordlund:2005a}, the coefficient
of the Spitzer heat conduction is reduced by a factor of three. The heat
conduction is the dominant process in the numerical scheme. Lowering the
conduction results in larger time steps, thereby allowing for longer time
series. Nevertheless, we use the coefficient as given in
\cite{Spitzer+Harm:1953} and \cite{Spitzer:1962}, i.e. three times more than
in \cite{Gudiksen+Nordlund:2005a}.

\begin{figure}
  \resizebox{\hsize}{!}{\includegraphics{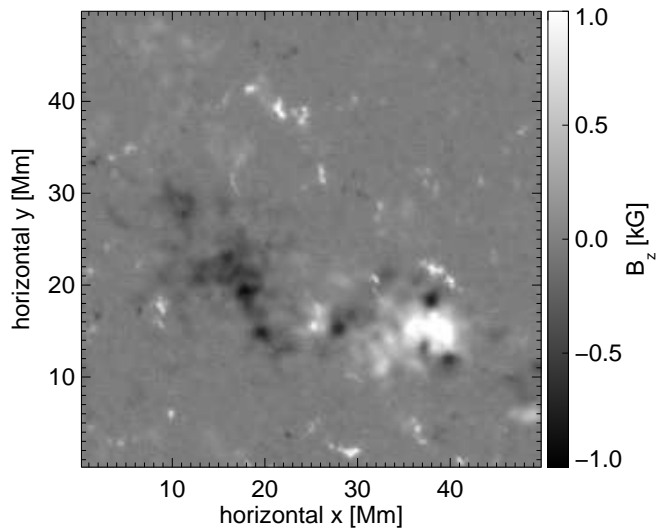}}
  \caption{Initial vertical magnetic field component at lower
    boundary.}
  \label{fig:mag_init}
\end{figure}

\subsection{MHD equations}
\label{sec:mhd-equations}%
The temporal evolution of the plasma and the magnetic field is governed by a
set of partial differential equations. These magneto hydrodynamic (MHD)
equations are written in terms of the logarithmic temperature and
logarithmic density, along with using the vector potential for the induction
equation. The former is due to the huge variation from the photosphere to
the corona, whereas the vector potential assures a solenoidal magnetic field
at all times. The equations for the mass density $\rho$, the fluid velocity
$ \vec{u}$,  and the temperature $T$ read as
\begin{align}
&  \frac{D \ln \rho }{D t} +  \nabla \cdot \vec{u} =
  0 \,,\label{eq:mass_con} \\
 & \frac{D \vec{u}}{D t}  =  \frac{1}{\rho} \left(-\nabla p + \rho
    \vec{g} + \vec{j}\times\vec{B}+
    2\nu\nabla \circ \left(\rho \underline{S}
    \right) \right) \,, \label{eq:mome_con}\\
  \label{eq:3} &\frac{D \ln T }{D t} + \left( \gamma -1 \right) \nabla \cdot \vec{u}=
  \frac{1}{c_V \rho T} \left[ \eta\mu_0 \vec{j}^2 + 2\rho\nu \underline{S}^2  +
  \nabla \vec{q} +N \right]\,.
\end{align}
Because of gauge invariance, we can add the gradient of an arbitrary scalar
field $\phi$ to the vector potential  without changing the magnetic
field
\begin{equation}
  \vec{B} = \nabla \times \left(\vec{A} + \nabla\phi\right)\,.
\end{equation}
We choose the resistive gauge $ \phi = \eta \nabla\cdot \vec{A}$.
For constant $\eta$, the induction equation in terms of the vector
potential reads as
\begin{equation}
\frac{\partial \vec{A}}{\partial t} = \vec{u} \times \left( \nabla
   \times \vec{A}\right) + \eta \nabla^2  \vec{A}\,.\label{eq:induc}
\end{equation}
The resistive gauge leads to a diffusion of the vector potential
proportional to the resistivity $\eta$, which is
preferable in numerical simulations.
Additionally the equation of state for an ideal gas correlates the
temperature with the pressure,
\begin{equation}
  p = \frac{k_B}{\mu \,m_p}\rho T \quad.
\end{equation}
We use the convective time derivative $D/Dt=\partial/\partial t +
\vec{u} \cdot \nabla$ to simplify the equations. The adiabatic constant
$\gamma=c_p/c_V$ is $5/2$, the mean atomic weight $\mu$ is
$0.667$ and $k_B$ and $m_p$ denote the Boltzmann constant and the proton
mass. The gravitational acceleration is $g = 274$\,m\,s$^{-1}$, and the
viscous force is given by the gradient of the traceless rate of strain
tensor $\underline{S}$, which only depends on derivatives of the velocity.

The resistivity $\eta$ and dynamic viscosity $\nu$ are constant and chosen
so that the corresponding grid Reynolds numbers,
i.e. R$=u_{\text{rms}}\text{dx} \,\nu^{-1}$ and R$=u_{\text{rms}}\text{dx}
\,\eta^{-1}$, where dx is the grid spacing and $u_{\text{rms}}$ the rms
velocity, in the model are close to one. For our grid resolution of
several hundred of kilometers, we chose therefore $\eta=10^{10}$m$^2$/s
and $\nu=10^{11}$m$^2$/s.
We apply a Newton cooling to the lowermost part of the model to adjust the
temperature to follow a profile similar to the average model of
\cite{Vernazza+al:1981},
\begin{equation}
  N = \frac{\rho c_V }{\tau_{cool}}\left(T_0 -T\right)\,,
\end{equation}
where $T$ is the temperature, $T_0=T_0(z)$ is the initial temperature
profile, and the cooling time scale is given by
\begin{equation}
 \tau_{cool}=\tau_0\exp\left(-z\right/h) \,,
\end{equation}
where $z$ is the height in the box.
The coefficients $\tau_0=10^{-5}$\,s and $h=40$\,km are chosen such that
the influence on the temperature above 3\,Mm is several orders of magnitude
less than the other physical processes described in equation~\eqref{eq:3}.
Heat is transferred by anisotropic Spitzer heat conduction. The heat flux vector is
\begin{equation}
  \vec{q}=  K_0\left(\frac{T}{[\mathrm{K}]}\right)^{\frac{5}{2}}\vec{\hat{b}}\left(
    \vec{\hat{b}}  \cdot\nabla T\right) \,,\label{eq:5}
\end{equation}
where $K_0=10^{-11}$\,W\,(m\,K)$^{-1}$ is the Spitzer value and
$\vec{\hat{b}}$ the unit vector of the magnetic field.
For numerical stability we include mass diffusion on the lefthand side of
equation~\eqref{eq:mass_con} and isotropic heat flux proportional to $ |\nabla
\ln T|\nabla T$ into the heat flux vector in equation~\eqref{eq:5}.
\subsection{Initial conditions}

\cite{Gudiksen+Nordlund:2005a,Gudiksen+Nordlund:2005b} used an observed
magnetogram with an spatial extend of 250x250\,Mm$^2$. Because of the
requested grid size, the magnetogram was scaled down by roughly a factor of 5
to fit into the computational domain. A resolution of 400\,km is required to
resolve both the granular motion and the temperature gradient in the
transition region. This downscaling removes small-scale magnetic network
patches. We use the same magnetogram as in
\cite{Gudiksen+Nordlund:2005a,Gudiksen+Nordlund:2005b} but to investigate
the influence of the interaction between the active region and quiet Sun
network fields small-scale features have to be introduced. The network flux
is taken from a second set of observations and is enhanced by a factor of
five to increase the interaction with the active region.  The resulting
magnetogram represents a small active region with a spatial extend of
50x50\,Mm$^2$ and is depicted in Fig.~\ref{fig:mag_init}. To fill the
computational domain with magnetic flux, we extrapolated a potential field.
The resulting corona was dominated by a large-scale loop connecting the main
polarities of the active region.

The initial plane-parallel temperature and density stratification match
the model atmosphere by \cite{Vernazza+al:1981}. The atmosphere is at rest at
the beginning.

\begin{figure}
  \resizebox{\hsize}{!}{\includegraphics{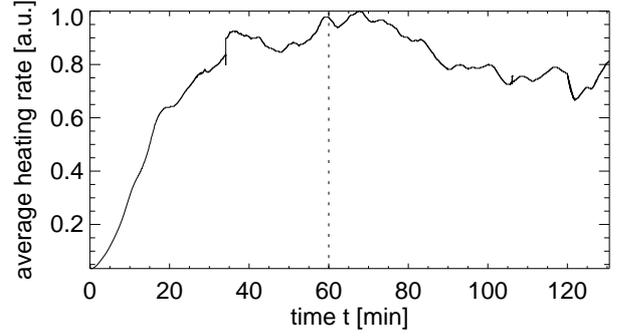}}
  \caption{Averaged heating rate over time starting at t=0, e.i. initial
    conditions at t=0. }
  \label{fig:j2m_t}
\end{figure}

Once the simulation is started, after some 30\,min solar time, the solution
is independent of the initial setup; e.g., the typical granule life time is
5\,min and the Alfv\`en crossing time about one minute.  Since we
start with a potential field, the averaged heating rate is zero at the
beginning. After initialization time, the heating rate reaches values that
are equal to the temporal average for the rest of the simulation
time. Figure~\ref{fig:j2m_t} shows the volume average of the squared current
density  beginning with the initial conditions. The dashed line in
fig.~\ref{fig:j2m_t} marks the time after which data was taken for the
analysis presented in this paper. This 
guaranties that data taken do not depend on the initial
condition. The atmosphere becomes highly structured and dynamic with high
velocities, but the overall and averaged appearance remains more or less
constant.

\subsection{Boundary conditions}

Photospheric motions at the lower boundary are the driving mechanism of the
dynamics in the upper atmosphere. These motions with power spectra similar
to observed velocity spectra shuffle around the foot points of magnetic
field lines. This leads to non force-free magnetic fields with currents
which heat the plasma by dissipation. The horizontal photospheric motions at
the lower boundary are computed in a similar way as done by
\cite{Gudiksen+Nordlund:2002,Gudiksen+Nordlund:2005a,Gudiksen+Nordlund:2005b}.
These random-like horizontal motions are time-dependent and prescribe a
velocity field similar to that found in the real solar photosphere from
observations. The vertical velocity at the bottom boundary, as well as all
three components of the velocity vector at the top boundary, are zero at all
times. The temperature $T$ and density $\rho$ at the bottom boundary are
kept at their initial values. At the top boundary, the temperature and
density have vanishing first derivatives in the vertical direction, but no
specific values for $T$ or $\rho$ are imposed.

The initial magnetic field configuration at the lower boundary evolves in
time following the induction equation. Because photospheric random motions
would corrode the active region, we update the magnetic field at the lower
boundary by its initial value. The time scale involved is chosen so that the
time series of the computed active region looks similar to an evolution of
an active region on the Sun. At the top boundary we assume a potential field.

In the horizontal directions the box is fully periodic.

\subsection{Numerical setup}
We use the Pencil Code \citep{2002CoPhC.147..471B} to run our numerical
model. It is a highly modular compressible MHD code tested on several
astrophysical problems and can be used for massive parallel computing using
Message Passing Interface (MPI). The numerical scheme is a finite difference
scheme comprising a sixth-order spatial derivative and a third-order Runge-Kutta
time-stepping scheme.
The calculation is performed using 128 grid points in each direction. This
is the minimum amount of grid points needed to resolve the granular motions on an
50x50\,Mm$^2$ area. The resulting grid spacing is on the order of
390\,km in the horizontal and 230\,km in the vertical direction,
respectively.

Parameters are set to their values found in literature except the
resistivity $\eta$ and the dynamic viscosity $\nu$. Both are set to fulfill
the requirement of a grid Reynolds number close to
unity.  The simulations runs for one solar hour before we start to collect
data with a cadence of 30 seconds for another solar hour. This gives us 120
snapshots of all physical variables such as velocity, temperature, density,
and magnetic field.

The model is conducted on a cluster consisting of 32 Intel CoreDuo (TM)
  Xeon processors. The cluster is located at the Kiepenheuer-Institut for
  Solar physics, Freiburg, Germany. The typical time step of the simulation
  is about one millisecond solar time, which results in an overall
  computing time on the order of 25.000 cpu hours.
\section{Results}
\label{sec:results}
\begin{figure}
  \resizebox{\hsize}{!}{\includegraphics{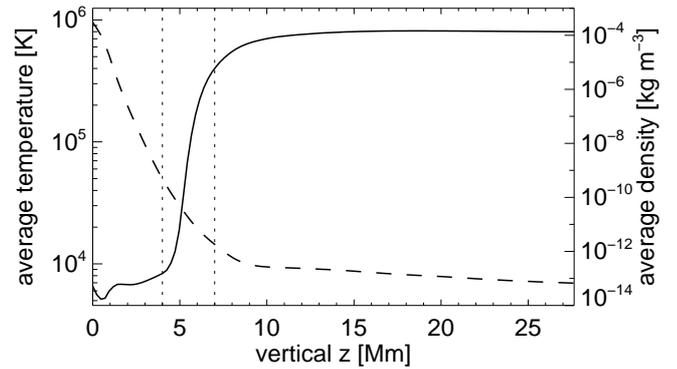}}
  \caption{Horizontally averaged temperature (solid) and density (dashed)
    over height for one snapshot. Vertical dotted lines mark the magnetic
    transition region (at 4\,Mm, cf. Sect.~\ref{sec:mag_trans}) and the
    maximum heating per particle (at 7\,Mm,
    cf. Sect.~\ref{sec:foot-point-dominated}).}
  \label{fig:h_averages}
\end{figure}

The model shows a highly dynamic and structured corona as a response to the
photospheric driving
 imposed at the lower boundary. The total energy is
balanced and the model corona reaches temperatures above one million Kelvin.
Horizontally averaged temperature and density profiles
(Fig.~\ref{fig:h_averages}) show the different layers starting from the
photosphere, the chromosphere, and the transition region followed by the
corona. The resulting average atmosphere is similar to a
\cite{Vernazza+al:1981} standard model. The density drops by several orders
of magnitude before it has a large constant scale height above 8\,Mm.  The
average coronal density is on the order of $10^{-13}$kg\,m$^{-3}$ or the
particle number density $10^{14}$m$^{-3}$.

\begin{figure}
  \resizebox{\hsize}{!}{\includegraphics{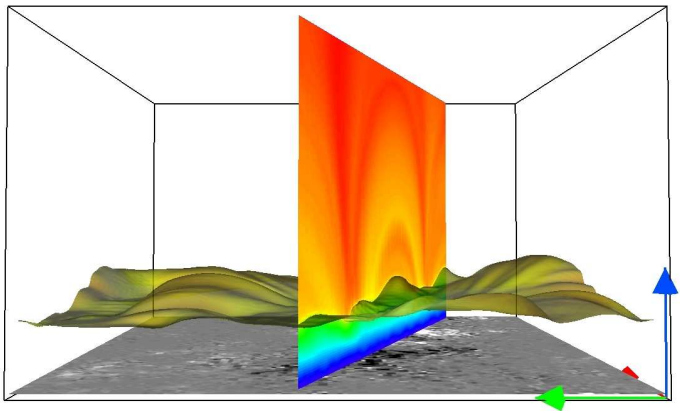}}
  \caption{Visible impression of transition region height. Isosurface of
  the temperature at log\,$T$/[K]=5.0 and a vertical cut through the domain showing
  the density above the main magnetic polarities (gray-scale bottom
  picture). Color code of the isosurface and the plane indicates logarithmic
  densities. A dense coronal loop is visible in the vertical cut. The domain
  shown is 50x50x30\,Mm.}
  \label{fig:transition}
\end{figure}

The transition region in the averaged temperature profile spans a height of
2\,Mm. Figure~\ref{fig:transition} shows the isosurface at 10$^5$K above the
vertical magnetic field component at the lower boundary. This highly
corrugated surface represents the center of the transition region.  The
density along the isosurface is not constant, and therefore the emissivities
of emission lines will vary.  The transition region above the main
polarities is at lower heights as the average transition region height. This is a
result of the anisotropic heat conduction. More heat is channeled along the
magnetic field lines towards the main polarities than to the network
flux because the corona is mostly connected to the strong magnetic flux
concentrations. Thus the transition region migrates downwards to regions of
higher density. The stronger heat income is then compensated for by increased
radiative loss (cf. Sect.~\ref{sec:model-corona}).

Figure ~\ref{fig:transition} shows
loop like structures with dense regions separated by loops with lower
density. The loops are filled and release their mass during the one-hour
simulation. The picture shows only one snapshot. A temporal analysis is done
in Sect.~\ref{sec:heating-rates-loop}.

In Fig.~\ref{fig:h_averages} the magnetic transition region is marked at a
height of 4\,Mm. This height depends on the typical length
scale of the magnetic features in the photosphere and is discussed in
Sect.~\ref{sec:mag_trans}. The height of the maximum of the
horizontally averaged heating rate per particle
(cf. Sect.~\ref{sec:foot-point-dominated}) is located at 7\,Mm (Fig.~\ref{fig:h_averages}).

\subsection{Average spatial distribution of the heating rate}

\begin{figure}
  \resizebox{\hsize}{!}{\includegraphics{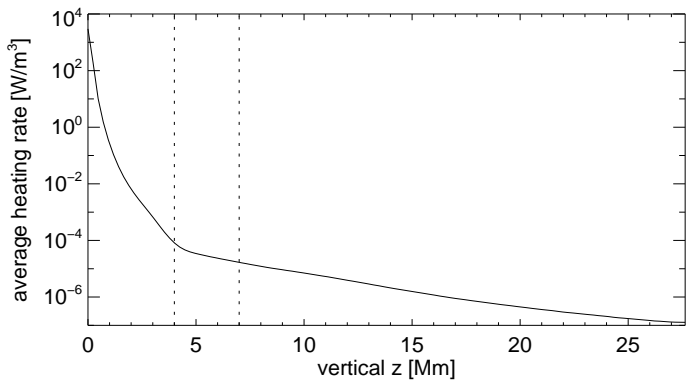}}
  \caption{Horizontally averaged heating rate per unit volume for one
    snapshot. In the upper part above $4\,$Mm the average heating rate drops
    exponentially with an almost constant scale height of about
    $5\,$Mm. Vertical dotted lines as in Fig.~\ref{fig:h_averages}.}
  \label{fig:heat_per_vol}
\end{figure}

The average heating rate drops dramatically with height up to some 4\,Mm and
then falls off more slowly but still exponentially (cf. Fig.~\ref{fig:heat_per_vol}). Below 4\,Mm
most of the small-scale magnetic field lines emerging from the photosphere
are closed back to the surface and thus causing the bend seen in
Fig.~\ref{fig:heat_per_vol}, which is discussed in
Sect.~\ref{sec:mag_trans}.
In contrast, the average heating rate per particle peaks at some 7 Mm height
(cf. Sect.~\ref{sec:foot-point-dominated}).
Overall, the average heating rate through Ohmic dissipation corresponds well
with the average Poynting flux into the upper atmosphere at all heights
(cf. Sect.~\ref{sec:energyfl-into-atmosp}).

\subsubsection{The magnetic transition region} \label{sec:mag_trans}

\begin{figure}
  \resizebox{\hsize}{!}{\includegraphics{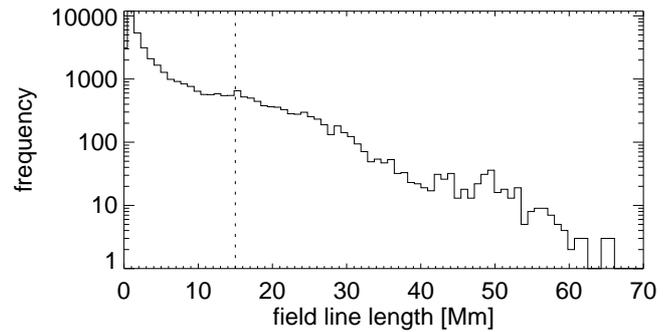}}
  \caption{\emph{Top panel}: Histogram shows the distribution of field line
    length in the domain respecting the periodic boundaries. Equally
    distributed at z=0 have been selected for the tracing algorithm 256$^2$
    starting points. Vertical dotted line depicts the local maximum at 15\,Mm. 
  }
  \label{fig:loop_stats}
\end{figure}

The vertical magnetic field at the lower boundary consists of two strong
opposite polarities and small-scale network flux patches. Field lines
emerging from the active region reach high up into the atmosphere and build
the coronal loop. Field lines starting at lower flux concentrations close
back earlier and thus are shorter. The top panel of
Fig.~\ref{fig:loop_stats} shows a histogram of lengths of field lines traced
from the lower boundary. The 256x256 starting points are equally distributed
in the x-y-plane. The number of field lines decreases in roughly inverse
proportion to length. But there is a change in the distribution function at
roughly 15\,Mm. Below, i.e. from 0 to 15\,Mm, the number of field lines per
length interval decrease more slowly than exponentially. The total fraction of
  magnetic field lines
in this interval is 75\%. These field lines connect network patches and
reach a height of roughly 4-5\,Mm when semi-circular loops are assumed. For
lengths above 15\,Mm the number of field lines decreases roughly
exponentially up to some 40\,Mm. For even longer field lines we find a
clustering with the longest field lines reaching almost the top of our
domain. These field lines connect the main polarities and are less
than 1\% in total.

A distinct change in the distribution of field line lengths is visible in
Fig.~\ref{fig:loop_stats} at a length of about 15\,Mm. This field line
length corresponds to an apex height of about 4\,Mm (assuming a
semi-circular loop).  This indicates that at heights in the computational
domain below roughly 4\,Mm the magnetic field topology is dominated by short
loops connecting small-scale features. Above this height the volume is
dominated by the longer field lines connecting the two main magnetic patches
of the active region. We therefore denote this height range at about 4\,Mm
where the magnetic topology changes from small-scale to large-scale as the 
\emph{magnetic transition region.} This can also be seen as a kink in the
distribution of currents (or more exactly the heating rate
${\propto}j^2$) with height in Fig.~\ref{fig:heat_per_vol}
(Sect.\,\ref{S:MTR}). This definition is similar to
\cite{2006A&A...460..901J}.

\subsubsection{Foot-point dominated heating} \label{sec:foot-point-dominated}
We analyzed the heating rate for each snapshot of our model by investigating
the Ohmic heating rate (cf. Eq.~\eqref{eq:3}) $\eta\mu_0\vec{j}^2$ at all
grid points. It corresponds to the volumetric heating rate and
Fig.~\ref{fig:heat_per_vol} shows its horizontal averages for one time
step. The volumetric heating rate decreases over more than ten orders of
magnitude from the photosphere to the corona.

From the photosphere to the upper chromosphere at 4\,Mm, the heating rate decreases
about eight orders of magnitude. Thus 99\% of the energy is deposited in the
chromosphere, and the chromosphere acts as an energy buffer. 

Above 4\,Mm (lefthand side in Fig.~\ref{fig:heat_per_vol}) the heating rate
drops exponentially with a scale height being on average around
5\,Mm. \cite{Gudiksen+Nordlund:2005b} found an average scale height of 5\,Mm
in their model, too. For both the transition region and the corona, the
heating scale height is constant. The volumetric heating rate at the top of
the chromosphere is approximately $10^{-5}$\,W\,m$^{-3}$.

\begin{figure}
  \resizebox{\hsize}{!}{\includegraphics{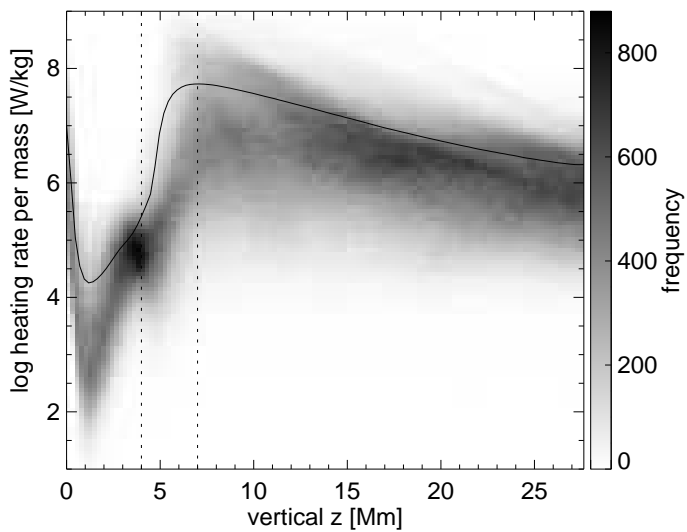}}
  \caption{Heating rate per unit mass. Scatter plot shows the distribution
    of logarithmic specific heating rate. Overplotted is logarithm of
    the horizontally averaged specific heating rate, i.e., heating per
    particle, for one snapshot. Vertical dotted lines as in
    Fig.~\ref{fig:h_averages}.}
  \label{fig:heat_per_mass}
\end{figure}

The density scale height (cf. Fig.~\ref{fig:h_averages}) at roughly 10$^4$K
(below 5\,Mm) is about 0.3\,Mm. Due the rapid temperature
increase up to $10^6$\,K the density scale height becomes 13\,Mm above
7\,Mm. Between the heights of 5\,Mm and 7\,Mm, the volumetric heating rate
drops exponentially with a scale height between the density scales at
these levels. This leads to a maximum specific heating rate per particle as
illustrated in Fig.~\ref{fig:heat_per_mass}.  The specific heating rate,
i.e. per particle, increases starting at the lower chromosphere with a scale
height of 0.5\,Mm, whereas the specific heating rate drops exponentially with
a scale height of 6\,Mm above 7\,Mm height. Thus, even though most of the
energy is deposited in the chromosphere, the available energy per particle is
higher in the corona.

The volumetric heating rate (cf. Fig.~\ref{fig:heat_per_vol}), as well as the
heating rate per particle (cf. Fig.~\ref{fig:heat_per_mass}), shows that the
heating of the coronal loops is foot-point dominated, and yet there is heating
also in the upper part of the corona.

\subsubsection{Energy flux into the atmosphere}
\label{sec:energyfl-into-atmosp}
\begin{figure}
  \resizebox{\hsize}{!}{\includegraphics{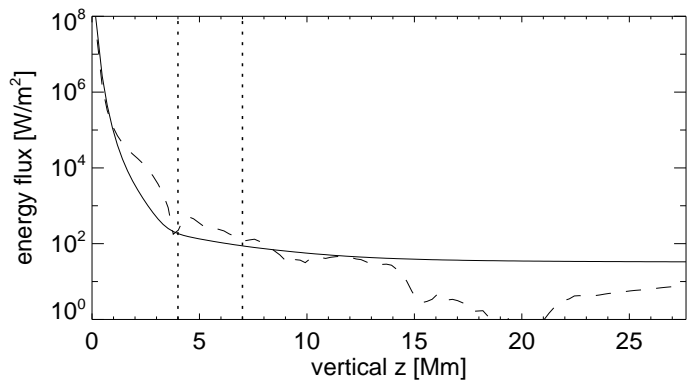}}
  \caption{Horizontal, averaged, upwardly directed heat flux (solid) derived
    from volumetric heating rate. Horizontally averaged absolute Poynting
    flux is overplotted as a dashed line. Both are derived from one snapshot
    in time. Vertical dotted lines as in Fig.~\ref{fig:h_averages}.}
  \label{fig:poyn_heat_f}
\end{figure}

The volumetric heating rate can be converted into an energy flux by assuming
that all energy comes from the lower boundary. This is the case for the
Ohmic heating as the driving occurs at the photosphere and the magnetic
field dominates the upper atmosphere.  The energy flux density $Q$ is a
function of height and gives a measure for the energy per unit time that has
to pass this height through a unit plane. The value of $Q$ can be derived
from the volumetric heating rate by integrating from $z$ to infinity;
i.e., the heating above $z$ is powered by the energy flux through the height
$z$,
\begin{equation}
  Q(x,y,z) = \int_{z}^{\infty} \eta \mu_0 \vec{j}(x,y,z')^2 \text{d}z' \;.
\end{equation}
The horizontal average of the energy flux density is depicted in
Fig.~\ref{fig:poyn_heat_f}. The energy flux into the upper
atmosphere is ranges from 10$^6$ to 10$^7$\,W\,m$^{-2}$ and decreases
rapidly up to 4\,Mm in height, the magnetic transition region. At the bottom
of the transition region the energy flux density is a few times
100\,W\,m$^{-2}$. This is consistent with the typical observation-based
estimations for the energy flux needed to heat and sustain the corona
\citep[e.g.][]{1977ARA&A..15..363W}. The energy flux decreases further to a
few times 10\,W\,m$^{-2}$ in the upper part of the corona.

The Poynting vector describes the direction and the strength of the energy
flux in an electro-magnetic field. Using the induction
equation~\eqref{eq:induc} and Ohm's law, we can write the Poynting vector
$\vec{S}$ independent of the electric field as
\begin{equation}
  \vec{S}=\frac{1}{\mu_0} \left( \eta \mu_0 \vec{j} - \vec{u}
    \times \vec{B} \right) \times \vec{B} \,.
\end{equation}
Figure~\ref{fig:poyn_heat_f} depicts the horizontally
averaged vertical component of the Poynting vector.  As the Ohmic heat is
energy converted from the magnetic field, the Poynting flux roughly follows
the energy flux derived from the heating rate. Because the Poynting flux depends
on the plasma velocities it shows more temporal
variation. \cite{Galsgaard+Nordlund:1996} found that the dissipation rate
scales linear with the Poynting flux at the lower boundary. Averaging over a
short time period we find that this relation is even valid for all
heights. The energy released in a sub-volume corresponds to the incoming
Poynting flux.

\subsection{Current sheets and nanoflares} \label{sec:curr-sheets-nanofl}
\begin{figure*}
  \sidecaption
  \includegraphics[width=12cm]{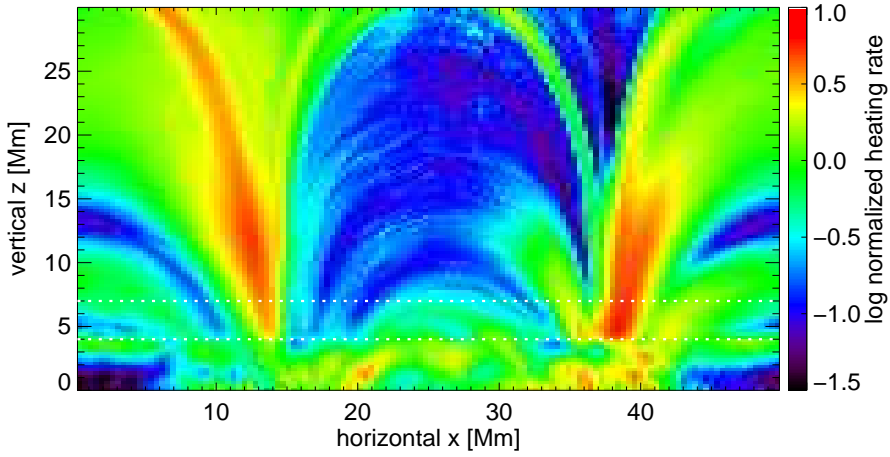}
  \caption{Integrated heating rate in a slab of the size $\Delta x$=50\,Mm,
    $\Delta y$=7\,Mm and $\Delta z$=30\,Mm. Color coded is the heating rate
    averaged along the y direction divided by its horizontal mean. The
    figure is periodic in the horizontal direction. Horizontal dotted lines
    as in Fig.~\ref{fig:h_averages}.}
  \label{fig:current_large}
\end{figure*}

The horizontally averaged heating rate in the simulation is almost constant
in time.  Investigating smaller volumes, i.e. at high spatial resolution,
the heating rate reveals a highly dynamic nature with a broad range of time
scales. Figure~\ref{fig:current_large} depicts
heating rate in a slab placed above the main polarities of the
magnetogram. The heating rate is normalized by its horizontal average in
order to remove the steep vertical gradient (cff Fig.~\ref{fig:heat_per_vol}).

\begin{figure*}
  \sidecaption
  \includegraphics[width=12cm]{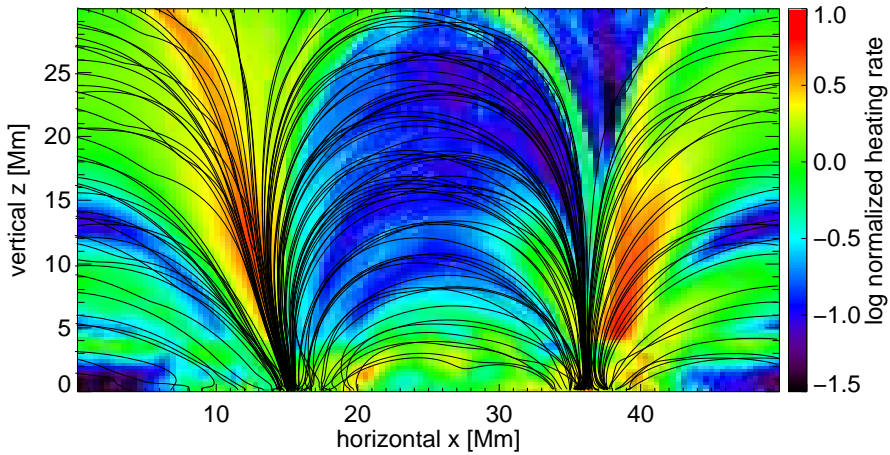}
  \caption{Integrated heating rate as in Fig.~\ref{fig:current_large}, but
    overplotted are the magnetic field lines in the slab projected onto the
    x-z plane. Starting points for the tracing algorithm are equally
    distributed in the 3D slab.}
  \label{fig:current_bf}
\end{figure*}

We used a narrow 3D volume, i.e. a slab, rather than a 2D
plane and integrate over one direction. A vertical cut through the domain
would only show the intersection of the magnetic field lines with the
plane but not a part of the loop. Samples of field lines that fit
completely into the slab are shown in Fig.~\ref{fig:current_bf}. The field
lines seem to intersect, which is only an effect of the projection onto the
x-z plane.  

In the coronal part the Ohmic heating rate is organized in current sheets
oriented parallel to magnetic field lines. These structures exist at the
resolution limit and are several grid cells wide. The alignment of the heating
structures to the field lines is illustrated in Fig.~\ref{fig:current_bf}.
Below 4\,Mm, the magnetic transition region, the normalized heating rate
shows small structures that indicate the short field lines
(cf. Sect.~\ref{sec:mag_trans}).

\begin{figure*}
  \sidecaption
  \includegraphics[width=12cm]{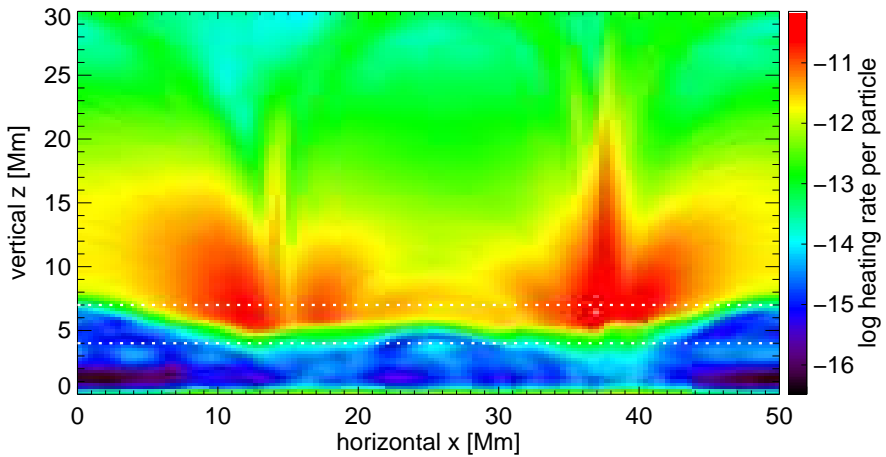}
  \caption{Logarithmic heating rate per particle in the same slab as in
    Fig.~\ref{fig:current_large} averaged in the y-direction. Horizontal
    dotted lines as in Fig.~\ref{fig:h_averages}.}
  \label{fig:current_heat}
\end{figure*}

The normalized Ohmic heating rate in Fig.~\ref{fig:current_large} only
illustrates the spatial distribution of the heating but does not give any
measures.  The specific heating rate per particle in the same slab is
depicted in Fig.~\ref{fig:current_heat}. It also shows the concentration of
the heating in structures along the magnetic field lines. Furthermore, the
foot-point dominated heating is illustrated. As shown in
Fig.~\ref{fig:heat_per_mass} the specific heating rate peaks at around
7\,Mm.

Current sheets along magnetic loops have also been investigate by e.g.
\cite{Rappazzo+al:2007} in high resolution loop simulations investigating
the Parker field line tangling that leads to Ohmic dissipation. Their model
compares well to our setup including constant resistivity, whereas
they used a so-called hyperresistivity. But in contrast our model extends
over larger volume and comprises a wide range of magnetic structures in a
more realistic geometry.

\subsection{Heating dynamics and response along field lines}
\label{sec:heating-rates-loop}
\begin{figure}
  \resizebox{\hsize}{!}{\includegraphics{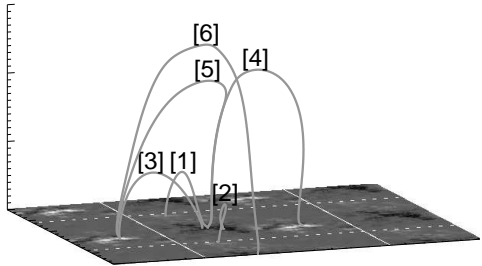}}
  \caption{3D representation of six magnetic field lines used for further
    analysis. White lines indicate the boundaries for the domain periodic in
    the horizontal directions. The box size is therefore 100x100x30\,Mm.}
  \label{fig:loop_3d}
\end{figure}

In the previous section we showed that the volumetric heating rate is
organized in current sheets. A closer look at the current sheets reveals
that these current sheet are discontinuous. Furthermore, the discontinuities
change in time. Therefore we analyzed the time dependence of the heating
rate along magnetic field lines. Figure~\ref{fig:loop_3d} illustrates the
selected six different field lines that are traced in the complete 3D box
respecting the periodicity of the domain. These field lines are selected
to represent different heights and different connectivities between the main
polarities, as well as into network patches. A list of properties of the
selected field lines is given in Table~\ref{ta:loop_props}. In the
subsequent discussion of the physical properties, we use the word loop
instead of the mathematical construct of a field line. Loop \#\,2 reaches
the bottom of the transition region, whereas loops \#\,1 and 3 extends above
the transition into the base of the corona. Loops \#\,4 to 6 extend into the
corona. In comparison to half circles, the apex height of loops \#\,1 to 3 is
smaller than half their foot-point distance, implying that these loops
appear somewhat flattened.  Loops \#\,4 to 6 are stretched out into the
corona with heights exceeding half the foot-point distance.
\begin{table}
\caption{Properties of loops shown in Fig.~\ref{fig:loop_3d}.}
\label{ta:loop_props}
\centering
\begin{tabular}{l l l l}
\hline
\hline
No & apex height  & length & foot point distance \\
   & [Mm] & [Mm] &  [Mm]\\
\hline
1 & 7.1  & 28.9  & 22.8 \\
2 & 5.0  & 26.5  & 22.9\\
3 & 8.9  & 33.8  & 25.2\\
4 & 24.7 & 76.7  & 25.0\\
5 & 23.7 & 76.5  & 26.7\\
6 & 29.0 & 87.5  & 48.9\\
\hline
\end{tabular}
\end{table}

For the following investigation we did not follow the field lines in time,
instead study the plasma properties in time on the trajectory defined by the
field line at the beginning of the time interval under study.  This is
justified by the low (e.i. some km/s) velocities perpendicular to the
magnetic field and the corresponding displacement over 10 minutes is close
to the grid spacing.

\subsubsection{Specific heating rate per particle}

\begin{figure*}
  \includegraphics[width=17cm]{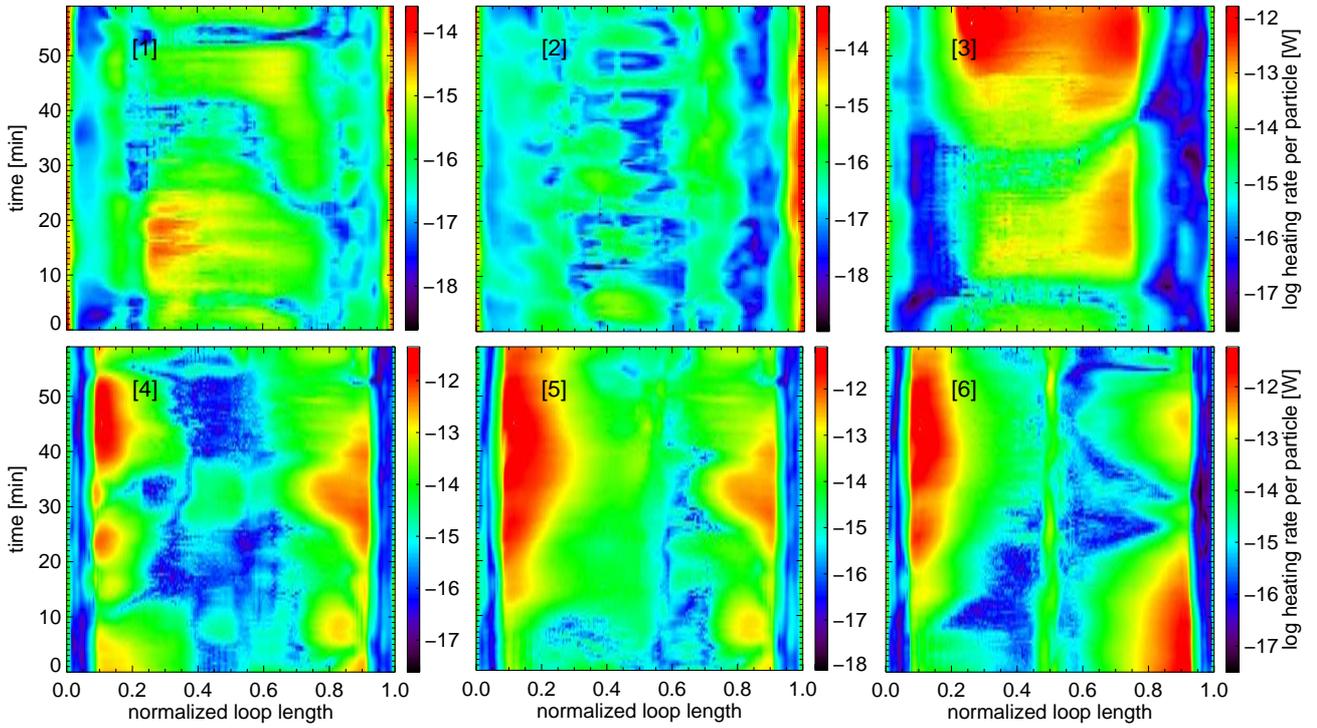}
  \caption{Heating rate per particle for one solar hour along  magnetic
    field lines depicted in Fig.~\ref{fig:loop_3d}. Loop lengths are given
    in Table~\ref{ta:loop_props}. The color table varies for each field line
    and is given on the righthand side of each panel. Loops \#\,1 to 3 are
    short, reaching any height below 10\,Mm, while loops \#\,4 to 6 represent
  coronal loops (see Fig.~\ref{fig:loop_3d}).}
  \label{fig:loop_time_heat}
\end{figure*}

Figure \ref{fig:loop_time_heat} displays the specific heating rate per
particle for one solar hour along the magnetic field lines. Different types
of heating events can be found. On the left side of panels [5] and [6], the
heating is continuous over more than 20\,minutes. On the other hand, loop
\#\,3 shows at the top (middle of panel [3]) short-lived heating
events. Since the numerical time step is only a few milliseconds these
events are resolved well in the simulation.

Loops \#\,1 and 2 connecting the main polarities with the network field
are mainly foot point heated at photospheric levels; however, the heating of
the foot points in the network (cf. righthand side of panels [1] and [2])
is stronger than the heating at the negative main polarity of active
region. As the velocities are quenched at strong magnetic fields, the shear
of the foot points becomes larger for the network flux patches.

Loop \#\,3 undergoes two main heating events at its top. The first event
(from 15\,min to 30\,min) consists of several short small-scale events. The
energy content of these small events can be computed and is similar to
nanoflares, i.e. some 10$^{17}$\,J, so that the first event compares to
nanoflare heating or heating by nanoflare storms. The second (starting at
30\,min) event starts with nanoflare heating before the heating gets
stronger. The entire top of loop \#\,3 is
heated in the last couple of minutes.

Loop \#\,4 is heated mainly at the height of the transition region where the
heating events have durations between 10 and 20\,min, and they occur on
both sides of the loop. Loops \#\,5 and 6 undergo a long heating event at
their foot points (on the lefthand side of panels [5] and [6]).

As already shown in Figs.~\ref{fig:heat_per_vol} and
\ref{fig:heat_per_mass} loops are predominantly heated at their foot
points. However, exceptions, e.g. loops \#\,1 or 3, can be found.

\subsubsection{Temperature variation}
\begin{figure*}
  \includegraphics[width=17cm]{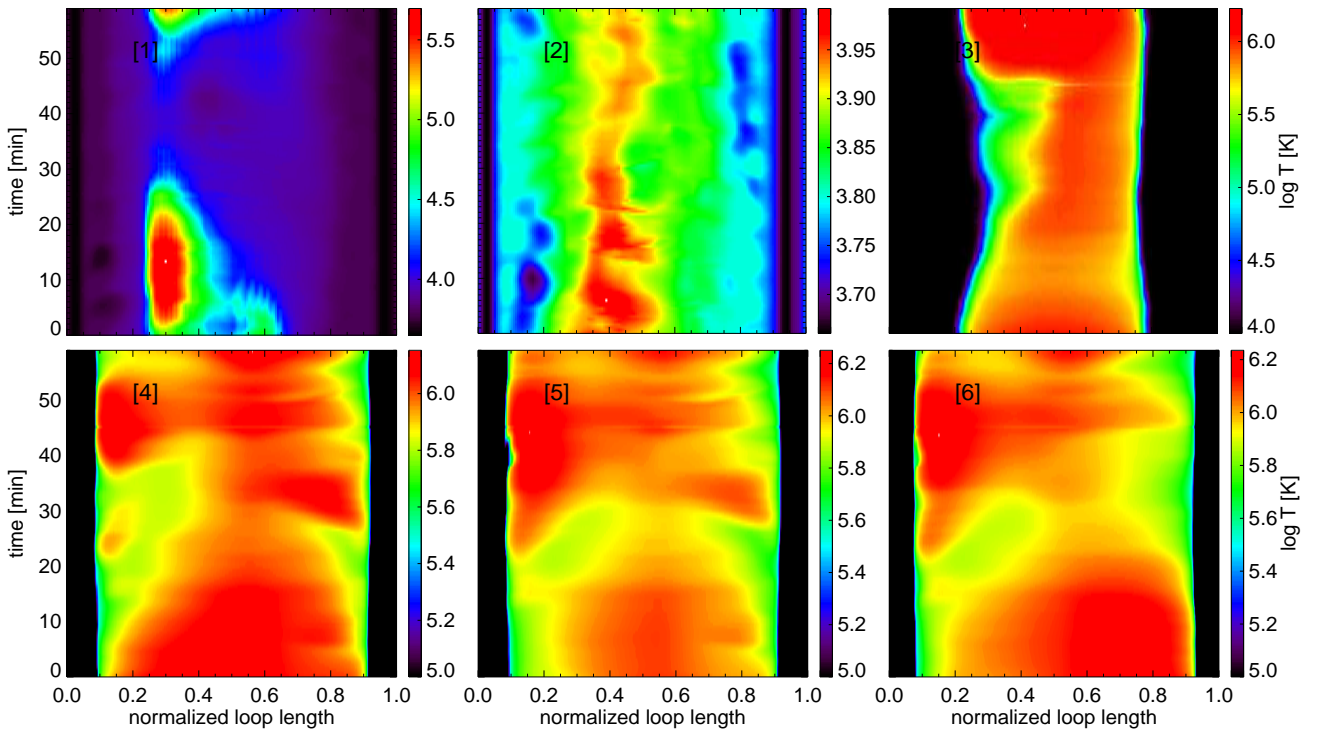}
  \caption{Same as in Fig.~\ref{fig:loop_time_heat}, but showing 
    logarithmic temperatures. The color table varies
    for each field line and is given on the righthand side of each panel.}
  \label{fig:loop_time3}
\end{figure*}

The temporal evolution of the temperature is depicted in
Fig.~\ref{fig:loop_time3}. Loop \#\,1 partly increases in temperature by an
order of a magnitude at the beginning of our time series. At this time also
the specific heating rate shows a peak.

Loop \#\,2 is the coolest and lowest loop. Its maximum temperature at the
top decreases with time, nevertheless, heating events seem to occur that are
not clearly visible in panel [2] in Fig.~\ref{fig:loop_time_heat}. Viscous
heating has to be taken into account to explain the small temperature
variations of this loop.  Because this loop is entirely embedded in the
chromosphere, the temperature (internal energy) is also slightly influenced by
the Newton cooling term (cf. Sect.~\ref{sec:mhd-equations}).

Loop \#\,3 shows a nice correlation between temperature variation and the
specific heating rate, but in comparison the temperature along the field
line is less structured. Due to the high efficiency of the anisotropic Spitzer
heat conduction, the internal energy is distributed along the field line on
short time scales. Small temperature variations due to the nanoflare heating
are smeared out rapidly.

Loops \#\,4 to 6 also illustrate  a nice correlation between the specific
heating rate and the temperature. At times when the foot points are not
heated the loop cools down. When the heating sets in not only small regions
but also the entire loop gets
hotter as a result of the efficient Spitzer heat conduction. For short time
periods the loops are almost isothermal.

\subsubsection{Vertical velocities}

\begin{figure*}
  \includegraphics[width=17cm]{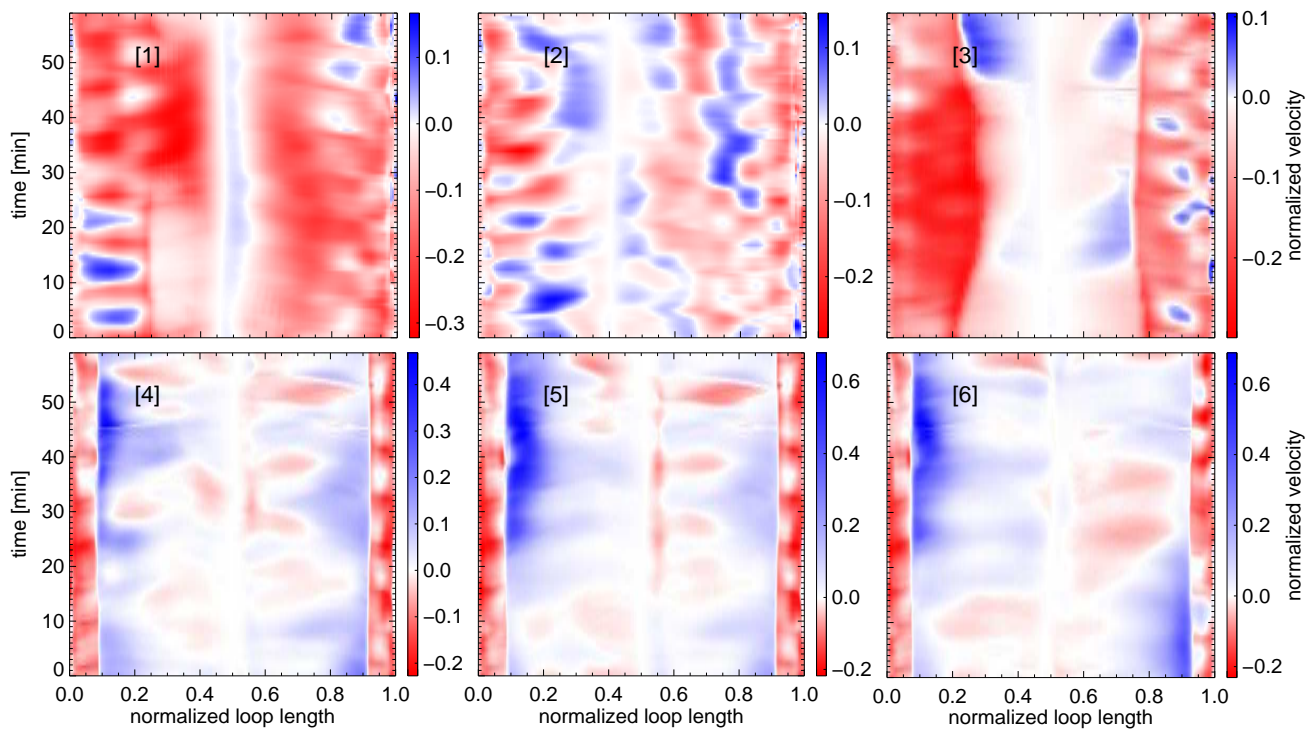}
  \caption{Same as in Fig.~\ref{fig:loop_time_heat}, but showing
    vertical component of flow velocity parallel to magnetic field
    lines. The velocity is divided by the
    local sound speed. The color table varies for each field line and is
    given on the right hand side of each panel. Red colors indicate
    down flows and blue colors up flows accordingly. In general, flows are subsonic.}
  \label{fig:loop_time_uu}
\end{figure*}

Figure \ref{fig:loop_time_uu} shows the temporal evolution of the vertical
component of the plasma velocity parallel to the magnetic field lines.
The velocity is divided by the local sound speed to resolve the large
difference between the photospheric values of a few
km\,s$^{-1}$ and the coronal flows  up to some 100\,km\,s$^{-1}$.
In general, the velocities in the box are subsonic.

By normalizing one can qualitatively distinguish between the types of motions
that are a response to the heating along the field line.  The signature of
the granular motions in the lower boundary is seen as periodic changes of up
and down flows at the foot points of the field lines. These motions have a
period of roughly 5\,min, which corresponds to the lifetime of the
granulation driving the magnetic field in the photosphere.

Loop \#\,1 is not heated much, and it cools down for the given time series. 
One effect is that the loop drains. The apex seems to move upwards, which can be
understood as a motion of the field line itself. As the motion is much
slower than one km\,s$^{-1}$, field line tracing can be neglected, as argued
above.

Loop \#\,2 shows a mix of up flow and down flow events along the loop. A
clear relation cannot be found when this is compared to the specific heating rates.
The chaotic like heating events result in quite irregular flow patterns.

Loops \#\,3 to 6 again show a nice correlation between the specific heating
rate (Fig.~\ref{fig:loop_time_heat}) and the vertical velocities
(Fig.~\ref{fig:loop_time_uu}). At places of strong heating the plasma
evaporates and is filled into the loop. The velocities are upwardly directed
(blue color) and ranges from  50\,km\,s$^{-1}$ to over
100\,km\,s$^{-1}$. At times of no heating, the coronal plasma cools down by
anisotropic heat conduction and radiative losses and starts to drain (red
color) out of the loop.

Loop \#3 reveals a strong downflow on the lefthand side. There the loop is
much denser than the opposite foot point. Thus the specific
heating per particle is less and the radiative cooling is more efficient.
The loop cools down quickly, and the plasma has to drain out of the loop.
After some 45\,min, the heating becomes strong enough so that plasma
evaporation can take place again.

Loop \#\,6 shows a siphon flow first from the right to the left side, and after
some 15\,min the siphon goes from the left to the right. These siphon flows
are a result of the asymmetric heating of the foot points. First the foot
point on the righthand side is heated, then the one opposite it. After some
45\,min, the loop is above one million degrees on both sides due to the heat
conduction. The draining stops and at both sides the loop is filled with
plasma.

Even though loops \#4 to 6 are foot-point heated either on one or the
other, a siphon flow is suppressed as soon as the loop approaches
isothermal. The highly efficient Spitzer heat conduction leads to the almost
constant temperatures for which the loops either are filled by plasma or
they drain.
\section{Discussion}
\label{sec:discussion}
Our approach of a 3D MHD  numerical model successfully produces
a one million degree hot corona above a small active region. The heating mechanism
following \cite{Parker:1978} self-consistently produces a heating rate strong
enough to balance Spitzer heat conduction and radiative loss.  The
empirically based photospheric motions create the dynamics of the system.
The model corona is in a quasi-stationary state where energy input by
the Poynting flux is balanced by radiative loss.  Other 3D models also find a
hot corona but use additional empirically based heating mechanisms
\citep[e.g.][]{2007ApJ...665.1469A}, or else the spatial extend is less
 \citep[e.g.][]{2009ApJ...701.1569M} than in the
model presented in this paper. In contrast to the models of
\cite{Gudiksen+Nordlund:2002,Gudiksen+Nordlund:2005a,Gudiksen+Nordlund:2005b},
 which are similar to ours, we also included the
magnetic flux of the network patches.

\subsection{Notes on the heating rate}

The heating rates found in our model are sufficient to sustain a hot corona.
Implemented Ohmic heating is only an approximation because the
magnetic resistivity used in the simulation is much greater than one would
expect for the coronal plasma following classical transport theory
\citep[e.g][]{Spitzer:1962}. The resistivity $\eta$ describes the process of
conversion of magnetic to thermal energy, and the value used in our model is
increased by about eight orders of magnitude, in order to have magnetic
grid Reynolds numbers of order unity. So that the currents are
actually dissipated.

The magnetic Reynolds number describes the
proportion of the advective to the resistive process for a certain length
scale or the proportion of the resistive time scale to the advective time
scale
\begin{equation}
  \textrm{Rm}=\frac{\vec{u}_{rms}l}{\eta}=\frac{\tau_{resis}}{\tau_{advec}}
\end{equation}
where $\vec{u}_{rms}$ is the rms velocity and $l$ a given length scale.
If we assume coronal values for the resistivity \citep[e.g.][]{Spitzer:1962},
then we have to look at cm length scales to obtain
a Reynolds number of 1 where the currents are dissipated. These length
scales are not resolved, but we must use the grid spacing to compute the
magnetic grid Reynolds number.  Using $l=$400\,km results in a
resistive time scale several orders of magnitude larger than the advective
time scale.  Thus we would not dissipate any currents in the time of our
simulation.  This contracts observations, beacause the life time of
active regions or eruptive events such as flares illustrates that the
resistive process is much faster. To circumvent this dilemma we chose the
resistivity $\eta $ such that the magnetic grid Reynolds number is
on the order of unity and therefore the time scales are
comparable. Naturally we find a wide spread of Reynolds numbers in our
domain.

\cite{Galsgaard+Nordlund:1996} also investigated the dependency of the
resistivity on the numerical resolution. For the first
event in their model they found a logarithmic scaling, but for the rest the
dissipation seems to be independent of resolution. Lowering the resistivity
would just postpone the dissipation process. As long as the Poynting flux
into the domain is constant, the dissipation rate is fixed on larger time
scales. \cite{Galsgaard+Nordlund:1996} find  a scaling law between the
dissipation rate and the Poynting flux that only depends on the granular
motions and the magnetic field in the photosphere. This
justifies our choice of $\eta$ to have a magnetic Reynolds number of
order unity.

\subsection{The magnetic transition region} \label{S:MTR}

At a height of about 4\,Mm the magnetic structure of the atmosphere is
abruptly changing (Sect.~\ref{sec:mag_trans}). This manifests itself most
clearly in the kink of the average heating rate and the energy flux as a
function of height (Figs.~\ref{fig:heat_per_vol} and
\ref{fig:poyn_heat_f}). This is related to the height where most of the
small-scale (granulation scale) magnetic flux closes, defining a
\emph{magnetic transition region}. Only above this height does the upper
atmosphere reach a magnetic state that represents the magnetic structure
of the corona.

This magnetic transition region is located above the height of the classical
canopy \citep{Giovanelli:1980,1992A&A...263..339S}, which is found below 1~Mm. The
latter is set by the rapid expansion of the magnetic field with height,
because the gas pressure drops exponentially and a horizontal equilibrium of
gas and magnetic pressure has to be achieved
\citep[e.g.][]{1990A&A...234..519S}. Basically the classical canopy is a
magneto-hydrostatic effect.

In contrast, the height of the magnetic transition region can be understood
by the potential field extrapolation of the distribution of magnetic flux at
the surface. In a numerical experiment \cite{2006A&A...460..901J} show that,
in quiet Sun regions above the network, the small-scale magnetic
concentrations would produce numerous short loops reaching up to about 4\,Mm
(cf.\ their Fig.~3). The small-scale structures push up the expanding field
lines from the larger magnetic patches, so in contrast to the classical
canopy, where the (predominantly horizontal) magnetic field is overlying a
field-free region \citep[e.g.][]{2000eaa..bookE2264S} a larger-scale
magnetic field is found above a volume with small-scale closed magnetic
field lines, in the case of the magnetic transition region .

Our current model carries this concept further by giving up the assumption
that the magnetic field is potential in nature (i.e.\ current-free). The
resulting currents found in the 3D MHD model are very strong in the
photosphere and chromosphere because of the shear applied by the horizontal
foot-point motions of the granular convection. The heating rate ${\eta}j^2$
drops roughly exponentially with a very small scale height of less than
0.5\,Mm.
Above the magnetic transition region, where the larger scale magnetic
structures dominate, the heating rate still drops exponentially (on
average), but now much more slowly with a ten times greater scale height of
about 5\,Mm.

\subsection{The chromosphere as an energy filter}

While the heating is highly intermittent in time and space, the horizontally
averaged heating rate is almost constant in time. At the magnetic transition
region, i.e. at the top of the chromosphere, the volumetric heating rate
drops to just below 10$^{-4}$~W\,m$^{-3}$. There the energy flux into the
corona heating up the upper atmosphere is a bit more than
100~W\,m$^{-2}$. This is consistent with typical estimates of the energy
demands derived from observations \citep[e.g.][]{1977ARA&A..15..363W},
especially when considering that we describe a \emph{small}
active region in our model. Our results are also consistent with
\cite{Gudiksen+Nordlund:2002,Gudiksen+Nordlund:2005a,Gudiksen+Nordlund:2005b}.

It is most remarkable that in our 3D MHD model the upwards directed energy
flux heating the corona is dropping by about six orders of magnitude
(Fig.~\ref{fig:poyn_heat_f}) from the surface to the base of the corona,
i.e. the magnetic transition region. This is consistent with the real Sun, because
in the model we have roughly the same Poynting flux at the surface as found
on the Sun. This implies that only the 10$^{-6}$th part of the energy flux
at the surface makes it into the corona. Or in other words: about 99.9999\%
of the energy is dissipated in the lower atmosphere below the magnetic
transition region! It has to be stressed that no fine-tuning was applied to
the model to get the correct energy flux into the corona, i.e. to
dissipate just 99.9999\% (and not all) of the energy in the lower
atmosphere.

Thus the lower atmosphere below the magnetic transition acts as an efficient
filter for the energy transported upwards to heat the corona. It would be 
very interesting to investigate how the efficiency of this filter changes
with the parameters of the 3D MHD model and the boundary conditions; i.e.,
what can be expected for other types of stars than our Sun. This could
affect how we understand why the X-ray luminosity in more active stars is
strongly enhanced \citep[up to 10$^4$;][]{2003A&A...397..147P}, while the filling factor of the coronal plasma can be increased by
a factor of only about 100 (assuming a filling factor for the Sun of about
1\%). Further studies will have to elucidate this problem.

\subsection{Individual strands and coronal loops}

While in a 1D numerical model for a coronal loop one can afford a high
spatial resolution to resolve strong gradients and shocks along the loops or
include ionization processes, a 3D MHD model provides the possibility of
investigating how neighboring structures interact and getting a better (more
self-consistent) description of the heating rate. Therefore it is 
interesting to compare our 3D model to state-of-the-art 1D loop models, as well
as to multi-stranded loop models.

\subsubsection{Thermal and dynamic response of individual loops}

If we investigate individual field lines in our 3D computational box, we can
follow the temporal evolution of the physical parameters in the same manner
as in 1D models for coronal loops (Sect.~\ref{sec:heating-rates-loop},
Figs.~\ref{fig:loop_time_heat} to \ref{fig:loop_time_uu}). This shows a
thermal and dynamic evolution of the loops, i.e. variations in temperature
and velocity, on time scales from well below minutes to one hour.

While the time scale of the driver in the photosphere is only
several minutes (granulation), the response of the corona in terms of the
heating rate shows much faster variations. This is a consequence of the
non linearity of the physical process of field line braiding and energy
conversion. Therefore the \emph{rapid} variations in temperature and velocity seen
in the coronal parts of individual loops are \emph{not} a signature of the
photospheric driving, but the response of the plasma to the small heating
events in rapid succession. In that sense the dynamics of the corona is
decoupled from the photosphere. (In the lowermost parts of the loops a clear
signal of the photospheric driver can be seen, of course,
cf. Fig.~\ref{fig:loop_time_uu}).

The variations on longer time scales are (partly) due to the long-term
evolution of the larger magnetic field patches in the photosphere. For
example, over a half hour the granular motions can create (or destroy) by
chance larger patches of stronger magnetic field, which in turn result in an
increased heating of the overlying corona (e.g.\ loops \#5 and \#6 in
Fig.~\ref{fig:loop_time_heat}). In general the evolution of the temperature
of a loop on these longer time scales roughly follows the heat input from below,
but with some modification because of the long cooling time of the coronal
plasma.

In general the dynamic evolution as found along field lines in our 3D models
compares well with state-of-the-art 1D loop models. For example, we find
dynamic processes such as cooling, draining, or siphon flows of the same
order of magnitude as comparable gradients as described, e.g., in
\cite{Mueller+al:2003,Mueller+al:2004}. While they assumed an exponential
drop of the heating rate in their loops with a comparable scale height, they
kept the heating constant in time. In the light of our finding that the
details of the photospheric driver (on short time scales) are not important
for the coronal dynamics, this can be considered as a minor difference.
Therefore we can conclude that our results for individual loops forming in
the 3D MHD model compare well to high-resolution 1D loop models. Of course,
the computational effort prevents us from including, e.g., ionization
processes, but we can properly resolve the thermal and the flow structure
within the loops. This allows us to also draw conclusions on multi-stranded
loop structures.

\subsubsection{Comparison to multi-stranded loop models}

In their multi-stranded loop models \cite{2006ApJ...647.1452P} assume that a
coronal loop consists of individual strands that are treated as independent
1D models. This is justified by the low heat conduction across the magnetic
field, which basically isolates the individual structures. Their strands are
heated by increasing the heating rate for some time (and by this mimicking
nanoflares).  In our 3D model a large loop, such as seen in the vertical cut
of Fig.~\ref{fig:transition}, contains numerous loop-like current sheets
that are parallel to the magnetic field, as depicted in
Figs.~\ref{fig:current_large} and \ref{fig:current_bf}. These current sheets
can be very thin, down to the size of the grid cells of the
computation. Thus one could consider a bundle of field lines with the
parallel current sheets (Fig.~\ref{fig:current_bf}) as the many strands
making up a loop in the multi-strand model.

However, there is a major difference to the multi-strand model. While the
cross-field conduction is very low, i.e.\ the strands are thermally
isolated, the 3D model incorporates interaction of the strands: the braiding
of the magnetic field lines causes the loop-shaped current sheets parallel
to the magnetic field. This results in a self-consistent impulsive heating
of the individual strands, whereas \cite{2006ApJ...647.1452P} release the
nanoflare energy ad hoc.  Recently 
\cite{2010ApJ...719..591L} have employed a
multi-strand model that mimics this process. They assume that the strands
are displaced by photospheric motions, which sets the time for the energy
release for a given strand according to \cite{Parker:1988}, but there
is still no interaction between the stands as these are described by 1D models.

Another difference to \cite{2006ApJ...647.1452P} is that they assume the
heating is distributed uniformly along the loop, while our model gives a
heating rate that drops exponentially with height. However, as
the \cite{2006ApJ...647.1452P} model loops are much larger than those fitting in our
computational box, it remains to be seen if this difference is significant
or if this is more a property of smaller than of larger ones.

More analysis of the distribution of the heating rate in space and time
as found in our 3D MHD model will have to show how this compares in detail
with the assumptions of the multi-stranded loop model, and might provide
important constraints for the further development of the multi-stranded loop
models.

\section{Conclusion}
\label{sec:conclusion}
Our 3D numerical model of the solar corona successfully describes how to
sustain a hot corona by the heating mechanism based on Ohmic
dissipation. The average heating rate and the derived energy flux
compare well to the observational requirements
\citep[e.g.][]{1977ARA&A..15..363W}. Furthermore, the model resolves the
intermittent and transient character of the heating on a wide range of
energy scales.

The heating rate per particle (or mass) is found to be strongest in the
transition region from the chromosphere to the corona. This is because of
the scale height of the (on average) exponentially dropping volumetric
heating rate is between the pressure scale heights of the chromosphere
and of the corona.

The heating is concentrated in current sheets that are roughly aligned with
the magnetic field lines and which are highly intermittent in time and
space. In general this supports the idea that the corona is heated by a
large number of small energy depositions, often named nanoflares
\cite[e.g.][]{Parker:1988}.

The dynamics within a corona loop (or a strand thereof), i.e. single field
lines in the complex 3D magnetic field, follow the transient heating
events. Loops with siphon flows, as well as irregular flow patterns, can be
found in the 3D MHD model.  The time-dependent heating rate along individual
field lines as found in our 3D MHD models may be used as input to higher
resolution 1D loop models including, e.g., non-equilibrium ionization.

The results show that 3D numerical box models of the corona are a useful
tool to investigate the nature of coronal heating, in particular the
distribution of the heating rate in time and space. The transient heating
can be investigated down to energies well below energetic events currently
observed on the Sun.

{ \acknowledgements We would like to thank Wolfgang Dobler for introducing
  us to the Pencil Code. Special thanks are due to Boris Gudiksen for the
  discussions when we started this project and for providing the code for
  the photospheric driver that we used in earlier experiments.  This work
  was supported through grants by the \emph{Deutsche Forschungsgemeinschaft}
  (DFG).
}

\bibliographystyle{aa}
\bibliography{literature,all,local}
\end{document}